\documentclass[epjST]{svjour}
\usepackage{graphicx}
\usepackage{url}
\usepackage{color}
\usepackage{listings}
\usepackage{dcolumn}
\usepackage{hyphenat}
\usepackage{amsmath}
\usepackage{amssymb}
\usepackage{cite}
\definecolor{darkred}{rgb}{0.75,0,0.25}
\definecolor{darkgreen}{rgb}{0,0.5,0.25}
\definecolor{purple}{rgb}{0.5,0,0.5}
\definecolor{grey}{rgb}{0.5,0.5,0.5}
\newcolumntype{t}[1]{D{.}{.}{#1}}
\newcolumntype{.}{D{.}{.}{-1}}

\begin{document}

\title{Random number generators for massively parallel simulations on GPU}
\author{Markus Manssen\inst{1}, Martin Weigel\inst{2,3}, and Alexander K. Hartmann\inst{1}}

\institute{Institute of Physics, University of Oldenburg, 26111 Oldenburg, Germany
  \and Applied Mathematics Research Centre, Coventry University, Coventry, CV1~5FB,
  United Kingdom \and Institut f\"ur Physik, Johannes Gutenberg-Universit\"at Mainz,
  Staudinger Weg 7, 55099 Mainz, Germany}

\abstract{High-performance streams of (pseudo) random numbers are crucial for the
  efficient implementation for countless stochastic algorithms, most importantly,
  Monte Carlo simulations and molecular dynamics simulations with stochastic
  thermostats. A number of implementations of random number generators has been
  discussed for GPU platforms before and some generators are even included in the
  CUDA supporting libraries.  Nevertheless, not all of these generators are well
  suited for highly parallel applications where each thread requires its own
  generator instance.  For this specific situation encountered, for instance, in
  simulations of lattice models, most of the high-quality generators with large
  states such as Mersenne twister cannot be used efficiently without substantial
  changes. We provide a broad review of existing CUDA variants of random-number
  generators and present the CUDA implementation of a new massively parallel
  high-quality, high-performance generator with a small memory load overhead.  }

\maketitle
\section{Introduction}

In the field of computer simulations \cite{hartmann:09}, the construction of suitably
good and fast pseudo-random number generators (RNGs) has been a long-standing problem
\cite{gentle:03}. This is mostly due to it being ill defined since, as John von
Neumann put it, ``anyone who considers arithmetical methods of producing random
digits is, of course, in a state of sin'' \cite{vonneumann:51}. In other words, true
random numbers cannot be produced from purely deterministic algorithms and hence the
degree to which the thus generated sequences of numbers resemble random sequences is
relative. A number of notoriously bad RNGs had been implemented as part of standard
libraries on the computer systems available in the early days of computer simulations
see, e.g., Ref.~\cite{marsaglia:03}. To separate the wheat from the chaff, a number
of collections or ``batteries'' of tests, comparing statistical properties of
pseudo-random sequences to those expected for a true random process, have been
suggested and extensively used in the past. While for many years Marsaglia's DIEHARD
suite \cite{marsaglia:diehard} was considered the gold standard in the field, with
the increase in computer power and the ensuing higher statistical precision of
simulations more stringent criteria have to be applied and so to today's standards
one would tend to require a generator to pass the SmallCrush, Crush and BigCrush test
series suggested by L'Ecuyer and coworkers in the framework of the TestU01 suite
\cite{lecuyer:07}.

Since, by definition, a pseudo RNG sequence is the result of a deterministic
algorithm, for each generator one can, in principle, construct tests that show that
the resulting sequences are not truly random. While the mentioned test batteries
check for general statistical flaws of the produced numbers, passing all tests does
not guarantee that a given generator does not lead to systematic deviations when used
in a Monte Carlo simulation of a specific system. In fact, such {\em application
  tests\/} serve as additional checks and have repeatedly helped to expose flaws in
generators \cite{ferrenberg:92,parisi:95}. Of particular value for such testing are
non-trivial models where exact solutions are available to allow for detecting
significant deviations without the need for simulations using other generators. A
most useful example in this respect is the two-dimensional Ising ferromagnet, for
which exact expressions for finite systems can be easily computed
\cite{ferdinand:69a}.

With the increasing interest in harvesting the superior parallel performance of
graphics processing units (GPUs) for general computational purposes \cite{owens:08},
these devices have also been discovered by researchers using Monte Carlo and
molecular dynamics simulations as convenient means of pushing the limits in studies
of notoriously difficult problems. Within a rather wide range of applications,
lattice spin systems with short-range interactions appear to benefit particularly
well from the massive parallelism of these devices
\cite{preis:09,yin:10,weigel:10c,ferrero:11,yavorskii:12}. In these systems, the
large-scale parallelism of threads and blocks is typically translated into parallel
updates of spins on non-interacting sub-lattices. A similar situation, only with
possibly a larger number of arithmetic operations interspersed with the consumption
of random numbers, is encountered for simulations of molecular and other off-lattice
systems \cite{meel:07,anderson:08}. To allow for efficient scaling to the large
numbers of threads required for the more and more powerful GPU devices available, the
central generation and distribution of random numbers by dedicated CPU or GPU threads
is not an option. Instead, each updating unit requires a distinct instance of a
RNG. Moreover, many calculations on GPU are rather limited by the bandwidth than by
the number and speed of available arithmetic units \cite{kirk:10}. To ensure good
performance, RNG related accesses to the device global memory must hence be kept at a
minimum. As a consequence, random number generators for such massively parallel GPU
simulations must fulfill requirements rather different from those for the
traditional, serial, CPU based simulations: (a) it must be possible to set up a large
number (thousands up to millions) of RNGs that deliver sufficiently uncorrelated
streams of random numbers and (b) to minimize memory transfers, the generator states
should be stored in local, shared memory, which is very limited.

The problem of parallel generation of random numbers is not completely new (see,
e.g., Ref.~\cite{brent:92}), but with the number of threads, say, in a multiple-GPU
simulation counting in the millions, it is brought to a new level. Different
strategies are conceivable here: (1) division of the stream of a long-period
generator into non-overlapping {\em sub-streams} to be produced and consumed by the
different threads of the application, (2) use of very large period generators such
that overlaps between the sequences of the different instances are improbable, if
each instance is seeded differently, or (3) setup of independent generators of the
same class of RNGs using different lags, multipliers, shifts etc. Note that even if a
given generator produces pseudo-random sequences of good quality (according to the
standard tests), a division into sub-sequences and the ensuing different order in
which the numbers are consumed, can lead to much stronger correlations and, hence,
worse quality than for the original generator (see the discussion in
Sec.~\ref{sec:lcg} below). A rather elegant solution to the problem of independent
generators was recently suggested in Ref.~\cite{salmon:11} and will be discussed
below in Sec.~\ref{sec:counter}.

The need to minimize memory transfers and hence ensure that the generator states can
be stored in shared memory seems to ask for RNGs with very small states. This appears
to rule out many of the generators popular for simulations on CPU. The Mersenne
twister \cite{matsumoto:98}, for instance, has a state of 624 words or about 20 kB
which, compared to the 48 kB of shared memory available to up to 1536 threads on an
NVIDIA Fermi GPU, is huge. As a rule of thumb, however, RNGs with a larger state lead
to larger periods and, in many cases, better statistical quality. To solve this
dilemma, two different strategies spring to mind: (1) the search for generators with
very small state, but good quality or (2) an attempt to {\em share\/} the state
between the threads of a single block and let these threads cooperate to generate
many random numbers from the same state in a vectorized call. As we will see below,
attempts to use the first approach with conventional generators generally do not lead
to satisfactory results. Notably, however, the concept of counter-based, stateless
generators of Ref.~\cite{salmon:11} discussed below seems to be an interesting
exception. The concept of state sharing is found, in general, to be more
successful. In Sec.~\ref{sec:xorshift} below, we discuss a new and efficient generator
designed along these lines, that passes all statistical tests. Another difference
between the CPU and GPU environments concerns the number of random numbers produced
and consumed on each invocation. While it is customary for many optimized CPU RNGs to
produce and store a significant number of entries in one (possibly vectorized) call
and store the produced numbers for later consumption, this is not feasible for GPU
generators with many threads consuming numbers in parallel and in the presence of the
memory limitations mentioned above.

Although research into RNGs suitable for GPUs is still at its beginning, a number of
such implementations has been discussed previously
\cite{curand,howes:07,meel:07,alerstam:08,preis:09,weigel:10c,weigel:10a,saito:10,demchik:11,salmon:11,nandapalan:11,bradley:11}. To
meet the design goal of a small memory footprint, we deliberately restrict our
discussion to generators using up to four machine words (128 bits) of state
information per thread. This appears to be about the upper limit for reaching good
performance on the present hardware for the massively parallel applications discussed
here. This rules out a number of implementations of general-purpose generators
\cite{howes:07}, such as XORWOW, a multiple recursive XORShift generator proposed by
Marsaglia \cite{marsaglia:03a}, implemented in the cuRAND library \cite{curand}
(which has 192 bits of state) as well as the standard Mersenne twister implementation
in the CUDA SDK (now superseded by MTGP \cite{saito:10}). We shortly discuss these
generators for completeness, however. For lack of space, we also here concentrate on
CUDA implementations and do not discuss RNGs on ATI cards which have slightly
different limitations \cite{demchik:11}. Unless stated otherwise, all test runs and
benchmarks have been performed with the CUDA 4.0 Toolkit. Finally, we do not here
consider the generation of quasi-random numbers. In Sects.~\ref{sec:lcg} and
\ref{sec:mwc} below, we focus on simple generators with small states, whereas in the
following Sects.\ (with the exception of Sec.~\ref{sec:counter}) the state-sharing
approach is discussed. All generators are benchmarked in terms of the quality of
random numbers by using the TestU01 suite, simulations of the 2D Ising ferromagnet as
an application test and GPU performance measurements.

\section{Linear-congruential generators\label{sec:lcg}}

Arguably the simplest and clearly the best understood RNGs \cite{knuth:vol2} are the
linear congruential generators of the form
\begin{equation}
  \label{eq:lcg}
  x_{n+1} = ax_n+c \pmod{m}.
\end{equation}
To convert the resulting sequence of integers to (uniformly distributed) numbers in
the interval $[0,1]$, one uses the simple output function $u_n = x_n/m$. If
appropriate constants $a$, $c$, and $m$ are chosen, the period of this class of
generators is $p = m$. Since for efficient implementations it is inconvenient (and
usually not computationally efficient) to make $m$ larger than the largest integer
representable in a native integer type, one is restricted to $m \le 2^{64}$ on
standard architectures. Hence, the achievable periods are rather small to today's
standards. What is more, on theoretical grounds it is argued that one actually should
not use more than $\sqrt{p}$ numbers out of such a sequence
\cite{knuth:vol2,gentle:03}. Indeed, for $m=2^{32}$ a simulation of a $4096\times
4096$ Ising system, for instance, would use $p$ numbers in only 256 sweeps. The
choices $m = 2^{32}$ or $m = 2^{64}$ have the advantage that there is no need to
perform the modulo operation explicitly since on most modern architectures (including
GPUs) integer overflows are equivalent to taking a modulo operation. For such power
of two moduli $m$, however, the period of the less significant bits is even shorter
than that of the more significant bits, such that the period of the $k$th least
significant bit is only $2^k$.

An advantage for the parallel calculations performed here is that one can easily skip
ahead in the sequence, observing that
\begin{equation}
  x_{n+t} = a_t x_n+c_t \pmod{m},
  \label{eq:lcg_skipping}
\end{equation}
where
\begin{equation}
  a_t = a^t \pmod{m},\quad
  c_t = \sum_{i=1}^t a^i c \pmod{m}.
\end{equation}
Therefore, choosing $t$ equal to the number of threads consuming random numbers, all
threads can generate numbers out of the same global sequence (\ref{eq:lcg})
concurrently. This corresponds to the sub-streams approach of parallel random number
generation mentioned above.  An alternative setup, that cannot guarantee the desired
independence of the sequences associated to individual RNG instances, however, starts
from randomized initial seeds for each generator, without using any skip-ahead
\cite{preis:09}. A potentially safer approach of choosing independent constants $a$
and $c$ for each instance while keeping $m$ unchanged does not appear to be feasible
since there are not enough multipliers with good properties available for massively
parallel simulations \cite{lecuyer:99}. For our tests with $m=2^{32}$, we used $a =
1\,664\,525$ and $c = 1\,013\,904\,223$, originally suggested in
Ref.~\cite{numrec}. Moving on to 64-bit, and hence increasing the memory footprint to
two machine words, gives a somewhat more reasonable, but still short period $p =
2^{64} \approx 2\times 10^{19}$. As multiplier we here chose $a =
2\,862\,933\,555\,777\,941\,757$ with provably relatively good properties
\cite{lecuyer:99}, where an odd offset, here $c = 1\,442\,695\,040\,888\,963\,407$,
needs to be chosen to reach the maximal period.

\begin{figure}[tb]
  \centering
  \includegraphics[keepaspectratio=true,scale=0.75,trim=45 48 75 78]{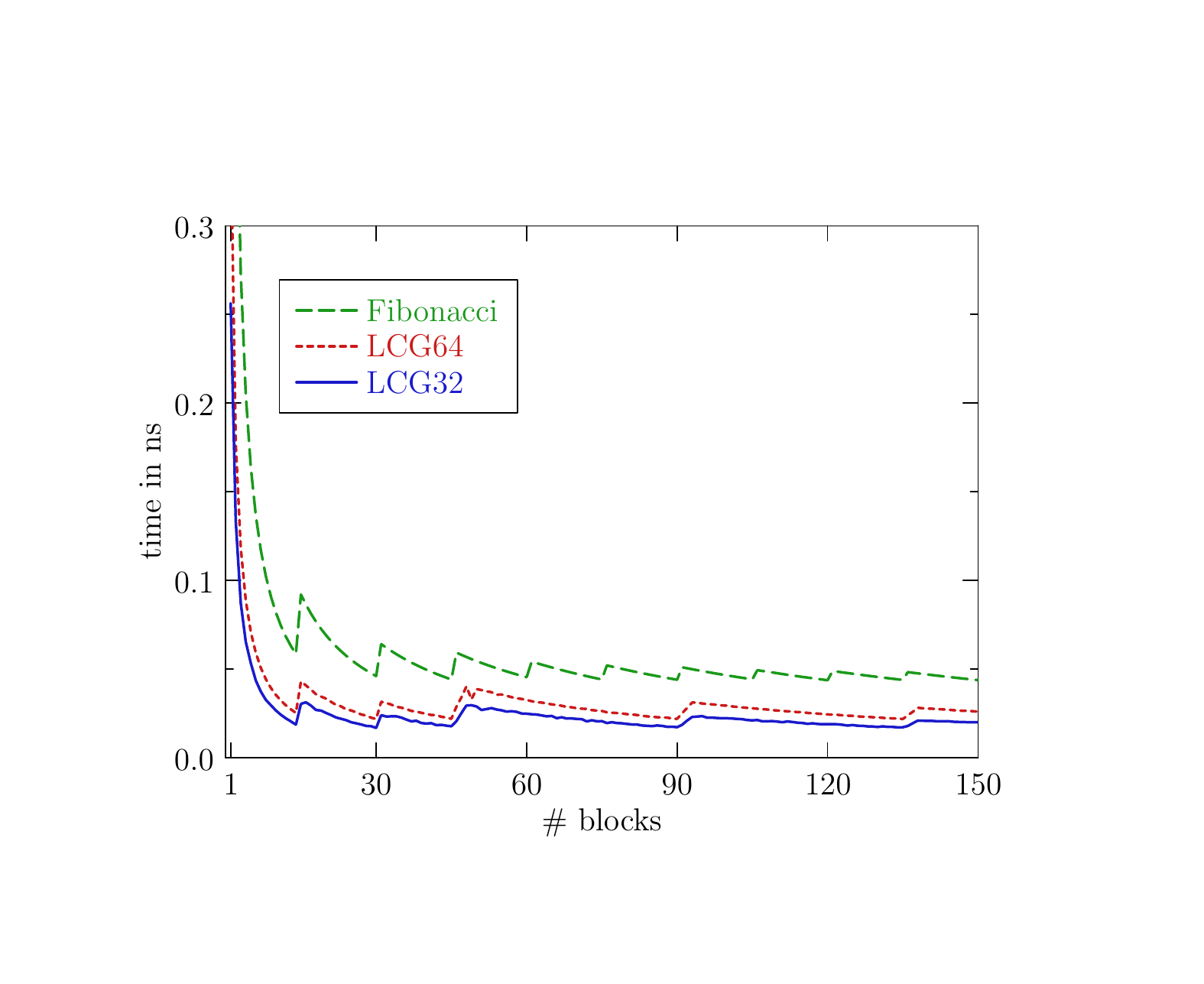}
  \caption{GPU time per random number for running different implementations of pseudo
    RNGs on a GTX~480 as a function of the number of grid blocks employed. The
    Fibonacci generator uses $r=521$ and $s=353$. All thread blocks have $512$
    threads.}
  \label{fig:rng}
\end{figure}

It is well-known that unmodified LCGs have rather bad statistical properties. In
particular, plotting $k$-tuples of successive (normalized) numbers as points in
$\mathbb{R}^k$, even for rather small $k$ the points are found to be confined to a
sequence of hyperplanes instead of being uniformly distributed. Applying the tests of
the TestU01 suite, we find that our 32-bit generator LCG32 already fails 12 out of
the 15 tests of the SmallCrush battery, such that we did not even attempt the Crush
and BigCrush tests. The results are summarized in Table \ref{tab:generators}. In a
parallel GPU code, this original order of using the produced numbers corresponds to
using skip-ahead according to Eq.~\eqref{eq:lcg_skipping}. The alternative approach
of randomly initializing each parallel instance of the generator with a different RNG
without any skipping provisions leads to somewhat better results as shown in the line
``LCG32, random'' in Table \ref{tab:generators}, such that most SmallCrush tests are
passed, but the generator fails 14 out of 144 tests in the Crush suite. The 64-bit
LCGs pass SmallCrush without failures, but have some problems in Crush. The
randomized variant of LCG64, in particular, tested with a number of 262144 threads
(as would be required in a simulation of a $1024\times 1024$ Ising ferromagnet
discussed below), fails a moderate 5 out of 196 tests in BigCrush (with an additional
3 suspicious results also counted as failures in Table \ref{tab:generators}) and
could thus almost be considered acceptable.

In addition to the generic tests provided by TestU01, we also performed an
application test in the form of a Metropolis simulation of a zero-field,
nearest-neighbor Ising ferromagnet on the square lattice. The GPU simulation code
employed has been discussed in detail in Refs.~\cite{weigel:10c,weigel:10a}. It uses
a double checkerboard decomposition which, if mimicked in a serial code, would result
in a specific order of using the random numbers. While this is only one particular
example problem, its discrete nature makes it relatively likely for flaws in the used
RNGs to show up in statistically significant deviations already of basic
quantities. While some deviations have been specifically observed for cluster updates
\cite{ferrenberg:92}, which we do not consider here (see, however,
Ref.~\cite{weigel:10b}), one could argue that the Ising model is amongst the problems
in lattice spin systems that is most sensitive to RNG correlations. In particular,
disordered systems that have extra (preferably high-quality) randomness from
quenched parameters in the Hamiltonian or continuous-spin models with the resulting
continuous probability distributions, are less likely to be afflicted with problems
resulting from flawed RNGs. Due to the availability of exact results for the internal
energy per site $e$ and specific heat $C_V$ on finite lattices \cite{ferdinand:69a},
statistical significance of deviations can be easily detected. For the generators
discussed here, we used a simulation of a $1024\times 1024$ system with a total of
$10^7$ sweeps after equilibration to determine $e$ and $C_V$ at the inverse
temperature $\beta = 0.4$. As shown in Table \ref{tab:rng} the 32-bit LCG, in
particular, leads to highly significant deviations in this example. What is more,
these deviations are aggravated in the GPU implementation by it sampling from the
same random-number sequence via skipping, but using random numbers in a different
order. If, on the other hand, the ``LCG32, random'' variant is used, the deviations
go away. As could be expected from the TestU01 results, the deviations produced by
the 64-bit flavor are significantly smaller and, again, disappear for the variant
with random seeds on GPU.

While the quality of pseudo-randomness is questionable, this class of generators
excels, however, in terms of its peak performance on GPU. In Fig.~\ref{fig:rng}, we
show the average time for generating a normalized random number uniformly distributed
in $[0,1]$ on fully loading a GTX~480 device with generator threads. The
characteristic zig-zag pattern results from the number of blocks being more or less
commensurate with the number of available multiprocessors on the GPU (which is 15 for
the GTX~480 used for these tests). The peak performance for the LCG32 is around
$58\times 10^9$ random numbers per second, whereas LCG64 yields $46\times 10^9$
numbers per second. Note that for the latter generator, we only use the 32 most
significant bits and hence generate only one random number per 64-bit calculation; on
using the whole state to produce two 32-bit numbers, the nominal performance would
double (but, of course, the quality of the generated numbers would be reduced). These
performance measures depend on how many numbers are produced with one generator
before the next generator state needs to be loaded from global memory. Hence, while
the performance numbers given in Table \ref{tab:generators} are from compute-bound
calculations and thus reflect the complexity of the arithmetic calculations to
produce a new random number, in many practical applications only a few numbers are
produced before the state of a generator needs to be written back to global memory,
such that in these memory-bound cases run-times are rather dominated by the number of
fetches from global memory required to load the generator state. This second type of
performance measure is reflected in the speed of the Ising simulation as shown in
Table \ref{tab:rng}. Here, the results for updating the spins of a tile once ($k=1$)
correspond to the situation that only two random numbers are produced per state load
and save operation, whereas for the simulation with multi-hit updates ($k=100$), 200
random numbers are produced per access to global memory. The good performance and
ease of implementation are plausible reasons for this type of generators having been
used in a number of recent studies with model implementations of GPU simulations
\cite{meel:07,preis:09,weigel:10c,kelling:11}.

\begin{table}[tb]
  \centering
  \caption{Internal energy $e$ per spin and specific heat $C_V$ for a $1024\times 1024$
    Ising model with periodic boundary conditions at $\beta = 0.4$
    from simulations on CPU and on GPU
    using different random number generators. $\Delta_\mathrm{rel}$ denotes the deviation from
    the exact result relative to the estimated standard deviation.
    The columns $t_\mathrm{up}^{k=1}$ and $t_\mathrm{up}^{k=100}$ 
    show the average time per spin update in ns for $k=1$ single and 
    $k=100$ multi-hit updates, respectively (see text). This test uses
    about $10^{13}$ or $2^{43}$ random numbers.}
  \vspace*{2ex}
  \begin{tabular}{l@{\hspace{0.2cm}}t{9}t{2}t{7}t{2}@{\hspace{0.25cm}}t{4}@{\hspace{0.1cm}}t{4}} \hline
    \multicolumn{1}{c}{method} & \multicolumn{1}{c}{$e$} &
    \multicolumn{1}{c}{$\Delta_\mathrm{rel}$} &
    \multicolumn{1}{c}{$C_V$} &
    \multicolumn{1}{c}{$\Delta_\mathrm{rel}$} & \multicolumn{1}{c}{$t_\mathrm{up}^{k=1}$} &
    \multicolumn{1}{c}{$t_\mathrm{up}^{k=100}$} \\ \hline
    exact  & 1.106079207   & 0 & 0.8616983594 & 0 &&\\ \hline
    \multicolumn{7}{c}{sequential update (CPU)} \\ \hline
    LCG32              & 1.1060788(15)      & -0.26    & 0.83286(45) & -63.45 &&\\
    LCG64              & 1.1060801(17)      & 0.49     & 0.86102(60) & -1.14 &&\\
    Fibonacci, $r=512$ & 1.1060789(17)      & -0.18    & 0.86132(59) & -0.64 &&\\ \hline
    \multicolumn{7}{c}{checkerboard update (GPU)} \\ \hline
    LCG32              & 1.0944121(14)      & -8259.05 & 0.80316(48) & -121.05 & 0.2221 & 0.0402\\
    LCG32, random      & 1.1060775(18)      & -0.97    & 0.86175(56) &  0.09 & 0.2221 & 0.0402\\
    LCG64              & 1.1061058(19)      & 13.72    & 0.86179(67) &  0.14 & 0.2311 & 0.0471\\
    LCG64, random      & 1.1060803(18)      & 0.62     & 0.86215(63) &  0.71 & 0.2311 & 0.0471\\
    MWC, same $a$      & 1.1060800(18)      & 0.45     & 0.86161(60) & -0.15 & 0.2293 & 0.0435 \\
    MWC, different $a$ & 1.1060797(18)      & 0.28     & 0.86168(62) & -0.03 & 0.2336 & 0.0438 \\
    Fibonacci, $r=521$ & 1.1060890(15)      & 6.43     & 0.86099(66) & -1.09 & 0.2601 & 0.0661\\
    Fibonacci, $r=1279$& 1.1060800(19)      & 0.40     & 0.86084(53) & -1.64 & 0.2904 & 0.0700\\
    XORWOW (cuRAND)    & 1.1060654(15)      & -9.13    & 0.86167(65) & 0.04 & 0.7956 & 0.0576\\
    XORShift/Weyl      & 1.1060788(18)      & -0.23    & 0.86184(53) & 0.27  & 0.2613 & 0.0721\\
    Philox4x32\_7      & 1.1060778(18)      & -0.79    & 0.86109(65) & -0.93 & 0.2399 & 0.0523 \\
    Philox4x32\_10     & 1.1060777(17)      & -0.85    & 0.86188(61) & 0.30  & 0.2577 & 0.0622 \\ \hline  
  \end{tabular}
  \label{tab:rng}
\end{table}

\section{Multiply with carry\label{sec:mwc}}

Due to the less than optimal quality of random-number streams generated by LCGs, a
number of generalizations and improvements {\em without\/} increasing the size of the
state have been proposed. One well-known example is the multiply-with-carry approach
initially suggested by Marsaglia \cite{marsaglia:91}. In the simplest case, one
considers the sequence
\[
\begin{split}
  x_{n+1} &= ax_n +c_n \pmod{m},\\
  c_{n+1} &= \lfloor (ax_n+c_n)/m \rfloor.
\end{split}
\]
In other words, the additive term $c_n$ in the $(n+1)$st step is the {\em carry\/}
from the previous iteration, hence the name multiply-with-carry (MWC). As for the
LCGs, a particularly efficient generator results from taking $m$ to be a multiple of
the intrinsic word length. We consider $m = 2^{32}$ here, which allows to pack the
whole state $(x_n,c_n)$ in a 64-bit integer variable. For suitably chosen $a$, the
period is found to be $p=am-2$ which, for $a$ only slightly smaller than $m = 2^{32}$
comes close to the period $p = 2^{64}$ of the 64-bit LCG. A GPU implementation of
this generator was originally suggested in the CUDAMCML photon simulator package
\cite{alerstam:08} and was recently used for the Potts model simulations reported in
Ref.~\cite{ferrero:11}. To achieve the full period, one requires $am-1$ as well as
$(am-2)/2$ to be prime (such that $am-1$ is a {\em safe prime\/})
\cite{marsaglia:03}. A standard choice for a single generator instance is
$a=4\,294\,967\,118$. Since these conditions on $a$ are sufficiently simple, it is
relatively straightforward to produce a large number of multipliers for parallel
execution of RNGs with a comparable period and comparable quality of the generated
random number streams. (This is in contrast to LCGs, where the conditions for full
period are more complicated and only a few combinations of parameters produce streams
of acceptable quality \cite{knuth:vol2,gentle:03}.)

We have used the batteries from the TestU01 suite to judge the quality of the
pseudo-random sequence produced by a single MWC generator. As is summarized in Table
\ref{tab:generators} it fares marginally better than a 32-bit LCG but, in fact, worse
than the 64-bit LCG considered in the previous section. This is somewhat surprising
in view of the fact that this class of generators is often considered superior to
LCGs, and the space requirements for $(x_n,c_n)$ is the same as that for the 64-bit
LCG. The performance of the well-engineered GPU implementation of
Ref.~\cite{alerstam:08} for the pure random-number production comes in at only
slightly below that of the 64-bit LCG with around $44\times 10^9$ 32-bit random
numbers per second on a GTX~480.

As mentioned above, it is possible to systematically generate multipliers by finding
the largest safe prime $am-1$ with $a$ less than $2^{32}$ (which is just the
$a=4\,294\,967\,118$ above) and then systematically working one's way down from there
towards smaller multipliers. To check for primality one uses probabilistic tests such
as the Rabin-Miller algorithm with parameters that make the occurrence of false
positives sufficiently unlikely. There are more than $10^6$ such multipliers for
$m=2^{32}$, which should be sufficient for most applications. While the efficient
generation of large numbers of multipliers is possible using arbitrary-precision
libraries such as the GNU Multiple Precision Library, these multipliers need to be
transferred to and stored in the GPU main memory, as well as loaded by each
independent thread prior to generating random numbers. Hence, as far as memory
bandwidth is concerned, it is fair to say that the state of this generator is in fact
$64+32=96$ bits.

To complement the tests on pure random number generation we also studied the
Metropolis simulation of an Ising model as an application benchmark. The
corresponding results are collected in Table \ref{tab:rng}. Comparing the estimates
of $e$ and $C_V$ to the exact results, we find complete statistical consistency in
terms of the observed fluctuations. For comparison, we also show the results of using
the same multiplier (but different seeds) for each generator instance, which also
yields satisfactory results, but this setup is found to yield slightly better
performance since the cooperative load operation of the multiplier field is no longer
required. The application performance of this generator is very similar to that of
the 64-bit LCG, with a moderate overhead seen in the $k=1$ case for the loading of
the extra 32 bit multiplier.

The similar class of subtract-with-borrow algorithms \cite{marsaglia:91} is the basis
for the RANLUX generator popular in high-energy physics \cite{luescher:94}. There,
additional skipping in the random-number sequence is used to get rid of the
short-ranged correlations. An implementation of this generator on ATI cards was
presented in Ref.~\cite{demchik:11}. It cannot compete in performance, however, with
some of the good-quality generators discussed below. Another generalization of the
multiply-with-carry generator is possible in the form of a multi-term lagged
Fibonacci generator with additional carry. This could be implemented on GPU using
state sharing rather similarly to the case of the more standard lagged Fibonacci
generators discussed next.

\section{Lagged Fibonacci generators}

A large number of RNGs with bigger state can be written in the form of a generalized
lagged Fibonacci sequence with recursion
\begin{equation}
  \label{eq:fibonacci}
  x_{n} = a_1 x_{n-1} \otimes a_2 x_{n-2} \otimes \cdots \otimes a_k x_{n-k} \pmod{m}.
\end{equation}
Here, the operator $\otimes$ typically denotes one of the four operations addition
$+$, subtraction $-$, multiplication $\ast$ and bitwise XOR $\oplus$,
respectively. Since a history of at least $k$ steps of generated $x_n$ must be kept
in memory in order to perform the recursion, this class of generator effectively has
state size $32k$ bits (assuming 32 bit wide variables $x_n$). In the following, we
choose $m = 2^{32}$. While generators with $\otimes = +$ are sometimes also known as
multiple recursive RNGs, the choice $\otimes = \oplus$ often goes under the name of
Tausworthe or shift register generator \cite{marsaglia:03,gentle:03}. Here, we
concentrate on the most commonly considered case of generators with two terms, i.e.,
two non-zero multipliers $a_r$ and $a_s$ such that
\[
x_{n} = a_s x_{n-s} \otimes a_r x_{n-r} \pmod{m},
\]
which are reminiscent of the Fibonacci series $F_n = F_{n-1}+F_{n-2}$ that led to the
name of this class of generators. For an appropriate choice of the multipliers $a_s$
and $a_r$ (as well as the initial values) and assuming $r > s$, it is possible to
achieve periods of $p = 2^r-1$ for $\otimes = \oplus$, $p = 2^{31}(2^r-1)$ for
$\otimes = \pm$, and $p = 2^{29}(2^r-1)$ for $\otimes = \ast$, respectively
\cite{brent:92}. If the lag $r$ is big enough, the periods can be made astronomically
large. For $\otimes = +$ and $r=1279$ considered below, for instance, we have $p =
2^{31}(2^{1278}-1) \approx 10^{394}$. 

As an example of this class of generators discussed in Ref.~\cite{weigel:10a}, we
consider $\otimes = +$ and use an implementation that works directly on the output
variables $u_n\in [0,1]$ by using floating-point arithmetic as
\begin{equation}
  \label{eq:used_fibonacci}
  u_n = u_{n-r} + u_{n-s} \pmod{1}.
\end{equation}
For good quality one needs $r \gtrsim 100$, leading to relatively large storage
requirements, but here the generation of $s$ random numbers can be vectorized by the
$n$ threads of a block. Hence, the ring buffer of length $r+s$ 32-bit words is shared
among the threads of a block, leading to a state size of $(r+s)/n$ words per thread.
If one chooses $s$ only slightly larger than the number of threads per block (and $r$
not too much larger than $s$), only a few words per thread are consumed. A number of
good choices for the lags $r$ and $s$ are collected in Ref.~\cite{brent:92}. As
examples, we use here $r=521$, $s=353$ and $r=1279$, $s=861$, respectively. Since one
cannot have a large number of independent, ``good'' pairs $(r,s)$ resulting in a
reasonable state size per thread (assuming a constant number of threads dictated by
the application), division into sub-streams is the only possible strategy for
achieving independence between the streams generated by different blocks. To this
end, one can use a modified skipping procedure similar to that discussed in the
context of LCGs as outlined in Ref.~\cite{brent:92}.  In view of the astronomic
period, however, it appears safe to just seed the ring buffers of the generators for
different blocks with an independent RNG.

The quality of the thus generated streams of pseudo-random numbers crucially depends
on the choice of the lags $r$ and $s$. For the example generator
\eqref{eq:used_fibonacci}, both choices $r=521$ and $r=1279$ pass the SmallCrush
battery of tests, cf.\ Table \ref{tab:generators}. The generator with the smaller lag
fails two variants of the gap test in Crush, whereas the choice $r=1279$ passes all
tests apart from a suspicious result for a random walk test in Crush. In BigCrush,
another suspicious result for a random walk test as well as a failed variant of the
gap test are found; all other tests are passed. Hence, it is safe to say that from a
theoretical perspective these generators produce random streams of satisfactory
quality if only $r$ is chosen large enough. Regarding the Ising application test on
GPU using random seeding of the initial states of ring buffers for different thread
blocks, we find a significant deviation from the exact result for the internal energy
for the smaller lag $r=521$ and full consistency for the larger lag $r=1279$ for the
$1024^2$ system at $\beta = 0.4$ considered here, cf.\ the corresponding entries in
Table \ref{tab:rng}. The performance of the generator crucially depends on the number
of threads used per block as the number of words per thread $(r+s)/n$ decreases as $n
\le s$ approaches $s$. The performance results in Tables \ref{tab:rng} and
\ref{tab:generators} are for $n=512$ threads, which is realistic for the applications
considered, but somewhat unfavorable for the $r=1279$ generator. The pure RNG peak
performance for either choice of $r$ is around $23\times 10^9$ 32-bit numbers per
second for the GTX~480, about half of the performance of the 64-bit LCG.  While these
performances are virtually identical for $r=1279$ and $r=521$, we find that $r=1279$
significantly slows down the Ising code as compared to $r=521$, see the performance
data in Table \ref{tab:rng}. The performance of the stand-alone generator is also
illustrated in Fig.~\ref{fig:rng} as a function of the number of thread blocks
employed.

Improvements of the outlined scheme are easily conceivable. Three-term recurrences,
for instance, are known to generate significantly improved random-number streams
already at smaller choices of the lags \cite{brent:92}. Alternatively, one might
consider using {\em multiplicative\/} lagged Fibonacci generators which have been
shown to be Crush-resistant \cite{lecuyer:07}. A generator that has been widely used
in spin-model simulations, including the simulations carried out on the FPGAs of the
Janus special-purpose computer \cite{belletti:08a}, is the following recurrence
suggested by Parisi and Rapuano in Ref.~\cite{parisi:95},
\[
\begin{split}
  x_n &= x_{n-24} + x_{n-55},\\
  u_n &= (x_n \oplus x_{n-61})/m,
\end{split}
\]
with $m=2^{32}$. This corresponds to a lagged Fibonacci generator with an additional
shift-register step to improve the quality of the output. For this specific choice of
lags, however, we find that the quality of the generated random-numbers streams is
relatively poor with two suspicious tests in SmallCrush and 8 failed tests in
Crush. In view of these results and the fact that this specific choice of lags does
not appear very suitable for vectorization with the number of cores available on the
GPUs considered here, we have not attempted a GPU implementation of this specific
generator. Another popular generator, suggested in Ref.~\cite{marsaglia:90} and
dubbed RANMAR, consists of the combination of a two-term lagged Fibonacci and a
second, arithmetic sequence. While good at its time, it fails a number of tests
already in smaller suites (partially due to the low resolution of 24 bits) and is
hence probably not appropriate any more for today's applications. A GPU
implementation of RANMAR on ATI cards has been discussed in Ref.~\cite{demchik:11}.

\section{Mersenne twister}

The very popular Mersenne twister generator \cite{matsumoto:98} is based on a variant
of the general lagged Fibonacci generator concept outlined in the previous section in
the form of a twisted generalized feedback shift register generator. For the case of
32-bit numbers, one chooses a Mersenne prime $2^k-1$ and sets $N=\lceil
k/32\rceil$. The generator is then based on the recursion
\begin{equation}
  \label{eq:mersenne_twister}
  x_n = (x_{n-N}|x_{n-N+1})A \oplus x_{n-N+M}.
\end{equation}
Here, $1 < M < N$ is the additional, smaller lag and $(x_{n-N}|x_{n-N+1})$ denotes
the concatenation of the $32-r$ most significant bits of $x_{n-N}$ with the $r$ least
significant bits of $x_{n-N+1}$, where $r=32N-k$. The $32\times 32$ bit-matrix $A$
defines the twist operation which is chosen in a specifically simple form that allows
for an efficient implementation in terms of shift operations. To improve the
equidistribution properties, the sequence $x_n$ is subjected to an additional
tempering transformation, such that the output sequence finally is
\[
u_n = \lfloor x_nT \rfloor,
\]
where the tempering bit-matrix $T$ is a suitably chosen combination of single-term
shifts and binary-and operations, see Ref.~\cite{matsumoto:98} for details. It can be
shown that, if the corresponding characteristic polynomial is primitive, this class
of generators achieves the maximal period of $p = 2^k-1$ \cite{matsumoto:98}. The
most popular choice is MT19937 with $k=19937$, leading to $N=624$ and $r=31$ with the
additional lag $M=397$. The period is maximal with $p = 2^{19937}-1 \approx 4\times
10^{6001}$. While the quality of these generators is generally good, they
systematically fail tests in the Crush suites related to ${\mathbb F}_2$ linearity.

Without the use of state sharing, implementations of this type of generator lead to
very large states, such that they are not suitable for the type of simulations
discussed here. A smaller version of Mersenne twister with $k=607$ and hence state
size of $\lceil 607/32\rceil = 19$ 32-bit words using an independent generator
instance for each thread has been part of the NVIDIA CUDA SDK for some time. This is
still significantly in excess of the (arbitrary) cut-off of 4 words per thread
adopted here. Due to the reduced state size and period, this generator fails some
random-walk tests in addition to the routines based on ${\mathbb F}_2$ linearity
\cite{saito:10}. A variant of the Mersenne twister more suitable for GPUs has been
suggested in Ref.~\cite{saito:10}. It uses state sharing and a somewhat different
transformation particularly suitable for GPU characteristics. Vectorization is
performed along the same lines as outlined above for the Fibonacci generator,
generating $N-M$ numbers in one parallel sweep. The choice of the Mersenne prime
$2^{11213}-1$ favored in Ref.~\cite{saito:10} leads to $N=351$ which, ensuring $M<95$,
allows for 256 threads to generate numbers simultaneously. The resulting state size
is hence $351/256 < 2$ words per thread plus the overhead for the transformation
parameters common to each thread block. This adapted generator, dubbed MTGP, has been
included in the NVIDIA cuRAND library of random-number generators starting with
version 4.1 \cite{curand}.

The Mersenne twister generator can be run with parameter sets chosen such that
distinct instances have distinct irreducible characteristic polynomials of the
transition function, at least making it plausible that these sequences are
uncorrelated \cite{saito:10}. To this end, a 16-bit ID is used as an input to a
separate, number-theoretic code that searches for an appropriate parameter set. It
is, however, not guaranteed that such a set of parameters can be found. For MTGP, a
different approach with 32-bit IDs is used which, however, does not guarantee to
generate distinct characteristic polynomials \cite{saito:10}. While this setup, in
principle, allows for a large number of parallel instances, the parameter search is
found to be very time consuming such that, for instance, finding a single parameter
set for $k=11213$ can take up to an hour on current hardware \cite{saito:10}. As a
consequence, cuRAND comes with a built-in set of 64 (presumably) independent
sequences --- far too few for many of the applications we are discussing
here. Another inflexibility comes through the restriction to 256 threads per block
which cannot be easily changed.

We have only benchmarked the MTGP code here, finding that it fails, as expected,
those tests in BigCrush based on ${\mathbb F}_2$ linearity. The performance is found
to be at an acceptable, but not outstanding $18\times 10^9$ 32-bit random-number
samples per second. As the implementations available in CUDA 4.1 (which is still
incomplete) and Saito's website are limited to 200 parameter sets, while our Ising
test uses 2048 blocks, in view of the expensive parameter search mentioned above we
have not attempted the Ising test here.

\section{XORShift generators\label{sec:xorshift}}

Another class of generators proposed by Marsaglia \cite{marsaglia:03a} is based on
the observation that the XORShift operation, i.e., the binary XOR between a word and
a shifted version of itself, can be performed very fast on modern computers as it
does not involve integer addition, multiplication or division, and it leads to
high-quality pseudo-random sequences.  Representing a word of size $w$ as a vector
of bits $x=(x_1,\ldots,x_w) \in \{0,1\}^w$, a left shift can be expressed as a matrix
multiplication $(x_1, \ldots, x_W)L = (x_2, \ldots, x_W, 0)$ with the left shift
matrix
\begin{equation}
  L=
  \begin{pmatrix}
         0 &      0 & \cdots &      0 \\
         1 &      0 & \cdots &      0 \\
    \vdots & \ddots & \ddots & \vdots \\
         0 & \cdots &      1 &      0
  \end{pmatrix}.
\label{leftshiftmatrix}
\end{equation}
If we denote by $I$ the identity matrix, an XORShift by $a$ positions can be written
as $x(I \oplus L^a)$, which corresponds, in C programming language, to the expression
\lstinline!x ^ (x << a)!. The recursion suggested in Ref.~\cite{marsaglia:03a}
consists of three shifts,
\begin{equation}
  x_n = x_{n-1}(I \oplus L^a)
  (I \oplus R^b)(I \oplus L^c) =: x_{n-1} M,
  \label{threeshift}
\end{equation}
where $R = L^T$ denotes the right shift and a word-size $w$ of $32$ or $64$ bits is
used. For an appropriate choice of the shifts $a$, $b$ and $c$, namely when the
characteristic polynomial $P(z) = \det(M - zI)$ is primitive, these generators attain
maximal period $2^w-1$. While, for $w=32$ or $w=64$, this is still rather low,
similar transition matrices for $w=96$, $w=128$ and $w=160$ are also suggested
\cite{marsaglia:03a}. The combination of the $w=160$ bit flavor with a 32-bit Weyl
generator (see below) defines the XORWOW RNG implemented in the cuRAND library
\cite{curand}. While this has reasonably good properties (including a period of
$2^{192}-2^{32}$), the state of $192$ bits per thread is larger than desirable.
Panneton {\em et al.\/} \cite{panneton:05} criticized Marsaglia's generators for poor
quality and tried to amend them by including more than three distinct shift
operations.  Brent \cite{brent:07}\footnote{Also note the summary at
  \url{http://papercore.org/Brent2007}.} instead concentrated on finding good
parameter sets, for multiple recursive generators, i.e. $(a,b,c)$ for the above
generator, by the use of heuristics.

\begin{figure}[htpb]
 \centering
 \includegraphics[scale=0.55]{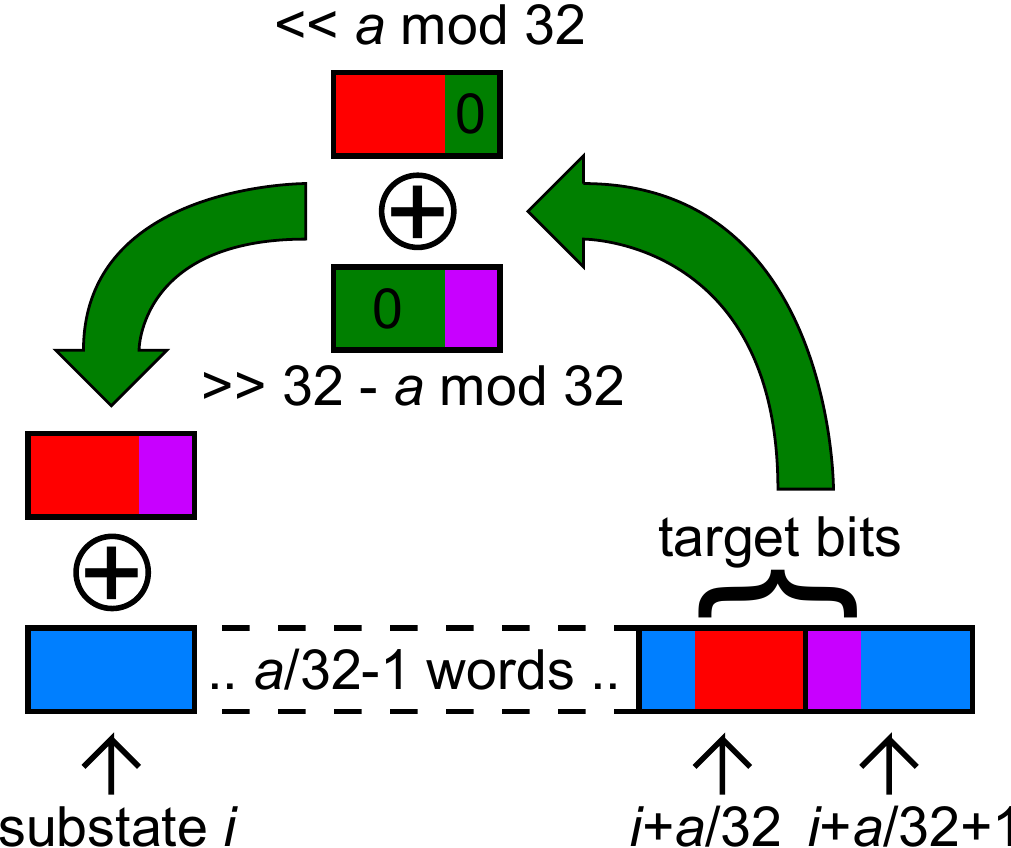} \hspace*{1.5cm}
 \includegraphics[scale=0.55]{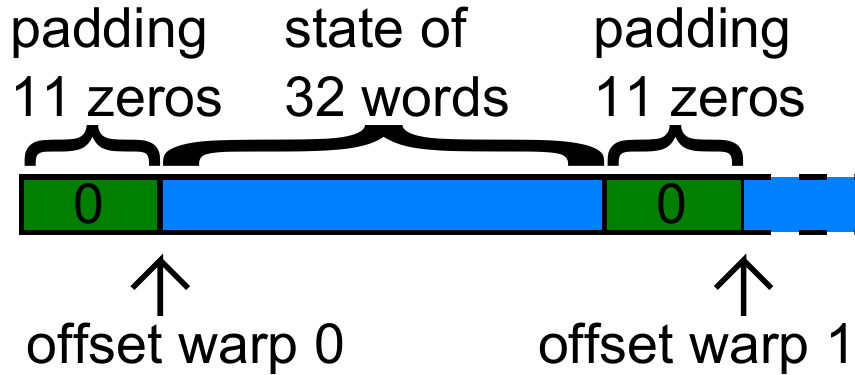}
 \caption{Left: left shift on a shared memory array, which stores the state of a
   RNG. For thread $i$ to update its part of the state, it needs a region of the
   state that spans the sub-states of threads $i+a/32$ and $i+a/32+1$. Right: Padding
   of the state array in shared memory. Zeros are placed in between the states of
   different warps to avoid the need for treating shifts crossing the state boundary
   separately.}
 \label{wordshift}
\end{figure} 

As a more space efficient approach than XORWOW with even better statistical
properties, we here suggest to use the three-shift generator (\ref{threeshift}), but
for a dramatically increased word size of $w=1024$ bits. Adopting Brent's heuristics
\cite{brent:07}, we performed a search for a generator with maximal period $2^{1024}
- 1$ using Sage \cite{sage} (and particularly the included library NTL
\cite{ntl}). We found a generator with the parameters $a = 329$, $b = 347$, and $c =
344$, which has a primitive characteristic polynomial of weight\footnote{The weight
  of a polynomial is the number of non-zero terms.} 475. This generator is
implemented on GPU by splitting the 1024 bits of state into a single
\lstinline!WORD!$=$32-bit word in each of the \lstinline!WARPSIZE!$=$32 threads of a
warp and using the threads of a warp to cooperatively update the 1024-bit state. The
XORShifts are then performed in the following way. Consider, for instance, a left
shift by $a$ bits as illustrated in the left panel of Fig.~\ref{wordshift}. The bits
arriving at the part of the state at thread $i$ originate from the most significant
bits of the word of thread $i+\lfloor a/32\rfloor$ and the least significant bits of
the word of thread $i+\lfloor a/32\rfloor+1$. These parts need to be shifted left and
right, respectively, to be assembled to the updated word at position $i$. Our full
CUDA implementation is shown in Listing \ref{kernelxorshift}. There,
\lstinline!NWARPS! denotes the number of warps per block and the type
\lstinline!state_t! is simply an unsigned integer.  This implementation benefits from
the fact that all three shifts correspond to a \lstinline!WORDSHIFT!$=\lfloor
a/32\rfloor$ of 10, leading to \lstinline!RAND_A! $= a \pmod{32} = 9$ and accordingly
\lstinline!RAND_B! $= 27$ and \lstinline!RAND_C! $= 24$.  Though shifts exceeding the
word-size could be taken care of using conditionals, we prefer to simply pad the
shared array with \lstinline!WORDSHIFT!$+1$ words of zeros as shown in the right
panel of Fig.~\ref{wordshift}. As the threads of a warp are always in sync, explicit
synchronization is not needed. However the shared array has to be marked as
\lstinline!volatile! to make sure the compiler writes all values to shared memory and
does not simply keep them in registers instead\footnote{Note that the very recently
  released version 4.2 of the CUDA toolkit allows for a direct exchange of data
  between threads within a warp using so-called ``warp shuffle'' functions. Using
  this feature for implementing the present generator would clearly free shared
  memory for other uses and, presumably, increase the overall performance.}.

\begin{lstlisting}[caption=Kernel XORShift., label=kernelxorshift]
/*
 * Updates the RNG state in cooperation with in-warp neighbors. 
 * Uses a block of shared memory of size 
 * (WARPSIZE + WORDSHIFT + 1) * NWARPS + WORDSHIFT + 1.
 * Parameters:
 * 	state: RNG state
 * 	tid: thread index in block
 * 	stateblock: shared memory block for states
 * Returns:
 * 	updated state
 */
__device__ state_t rng_update(state_t state, int tid,
                              volatile state_t* stateblock) 
{

/* Indices. */
int wid = tid / WARPSIZE; // Warp index in block
int lid = tid % WARPSIZE; // Thread index in warp
int woff = wid * (WARPSIZE + WORDSHIFT + 1) + WORDSHIFT + 1; 
                                                   // warp offset
/* Shifted indices. */
int lp = lid + WORDSHIFT; // Left word shift
int lm = lid - WORDSHIFT; // Right word shift

/* << A. */
stateblock[woff + lid] = state; // Share states
state ^= stateblock[woff + lp] << RAND_A; // Left part
state ^= stateblock[woff + lp + 1] >> WORD - RAND_A; // Right part

/* >> B. */
stateblock[woff + lid] = state; // Share states
state ^= stateblock[woff + lm - 1] << WORD - RAND_B; // Left part
state ^= stateblock[woff + lm] >> RAND_B; // Right part

/* << C. */
stateblock[woff + lid] = state; // Share states
state ^= stateblock[woff + lp] << RAND_C; // Left part
state ^= stateblock[woff + lp + 1] >> WORD - RAND_C; // Right part

return state;
}
\end{lstlisting}

In view of the large period of the generator, we use skip-ahead to partition the
sequence into substreams to be used by different warps.  For this purpose, we decided
to split the whole random number sequence into blocks of $2^{137} \approx 2\times
10^{41}$ numbers, assigning them successively to the instances.  With current
hardware, it appears impossible for any of the instances to exhaust their
sub-sequences. To this end, we let all warps start with the same state and have them
skip the appropriate number of update steps by multiplying the state with the
precomputed $2^{137}$-th power of the recursion matrix $M$. The matrix, included in a
header, is copied to the device and bound to a texture to perform a matrix
multiplication.  Starting from the resulting states after this initialization the
warps can simply continue to update normally as shown in the previous section. If a
large number of instances is seeded, we found that this approach leads to very long
kernel running times, which can lead to problems if the used GPU is concurrently used
for display purposes as well. Such problems can be avoided by including higher powers
of the block skip matrix to reduce the number of multiplications needed. An
alternative, more sloppy approach consists of simply seeding each generator instance
with another RNG which, in view of the large period, should also prevent any
overlapping of sequences for all practical purposes.

Similar to the approach suggested in Ref.~\cite{marsaglia:03a}, we also considered a
combination of the XORShift generator with a simple {\em Weyl sequence},
\begin{equation} 
  y_n = (y_{n-1} + c) \mod 2^w,
  \label{weylgen}
\end{equation} 
with an odd constant $c$. As for XORWOW, we chose $w = 32$ and $c = 362\,437$ and,
following Brent \cite{brent:07}, return $y^i_n(I \oplus R^\gamma) + x^i_n
\pmod{2^w}$, where the superscript $i$ refers to the local state of thread $i$ and
$\gamma = w/2$. The period of the resulting generator is thus increased to $(2^{1024}
- 1)2^{32}$. Brent argues that in the XORShift generator elements with low Hamming
weight (i.e., small numbers of non-zero bits) will be followed by other low weight
elements. For our generator with $w=1024$, the probability for such events is,
however, astronomically small. Still, since the extra cost of the outlined
combination with a Weyl sequence is relatively small, we include it for the
measurements and tests reported below.  The calculation of any state of the Weyl
generator is trivially possible in one step,
$$y_n = (y_0 + nc) \mod 2^w,$$
such that there is no need to store the state in memory in between
invocations\footnote{Note that the seed of the Weyl generator is the same for all
  instances since the chosen size $2^{137}$ of sub-streams is a multiple of the Weyl
  generator's period $2^{32}$.}.

To assess the statistical quality of the resulting random-number sequence, we
subjected it to the batteries in the TestU01 suite. These tests were performed for
two orders of the generated numbers, namely (a) {\em single-thread order\/}, feeding
the sequence generated by a single thread to the test, and (b) {\em warp order\/},
feeding the 32 numbers generated by each warp in one step sequentially to the
test. In both cases, we found that all tests were passed. In addition, we tested for
equidistribution directly.  If we take output values $u_n$ of our generator, we
expect them to fill the interval $[0, 1)$ uniformly. Thus if we divide the interval
into $2^l$ cells, each cell should be hit the same number of times. More generally,
vectors made of $t$ successive numbers should evenly fill the $[0, 1)^t$ hypercube.
We implemented these tests as described in Ref.~\cite{panneton:05}, finding
acceptable uniformity. As discussed by Panneton \cite{panneton:05}, there are
different choices of the matrix $M$ in Eq.~\eqref{threeshift} leading to the same
characteristic polynomial. Of the four possible choice, we found one with rather bad
equidistribution properties with the other three being acceptable, and our choice
being the best of the available options.

In terms of performance, we find the standalone generator to produce a good $18\times
10^9$ uniform 32-bit random-numbers per second\footnote{Somewhat better results are
  found for version 4.1 of the CUDA Toolkit.}. As expected from the statistical
testing, the Ising application test does not reveal any significant deviations from
the exact results, see the corresponding data in Table \ref{tab:rng}. The performance
in the single-hit and multi-hit versions of the Ising application is reasonably good,
benefiting from the small state and arithmetic simplicity of the generator. The
source code of this generator is included for future reference in the Supplementary
Material of this review article.

\begin{table}[tb]
  \centering
  \caption{Overview of GPU random-number generators discussed in this review. The
    memory footprint is measured in bits per thread. For the TestU01 results,
    if (too many) failures in SmallCrush are found, Crush and BigCrush are not attempted;
    likewise with failures in Crush. The performance column shows the peak number
    of 32-bit uniform floating-point random numbers produced per second on a fully loaded GTX~480
    device. Note that the Philox generators, albeit
    occupying local memory of $4\,\times\,32$ bits for number generation, do not require to
    transfer a ``state'' from and to global memory as long as the generator keys are
    deduced from intrinsic variables such as particle numbers etc.}
  \label{tab:generators}
  \begin{tabular}{l@{\hspace{0.2cm}}c@{\hspace{0.15cm}}c@{\hspace{0.15cm}}c@{\hspace{0.15cm}}ccl} \hline
    \multicolumn{1}{c}{generator} & bits/thread & \multicolumn{3}{c}{failures in TestU01} & Ising test & perf. \\
              &                  & SmallCrush & Crush  & BigCrush        &   & $\times 10^9$/s \\ \hline
    LCG32     &               32 &      12    &   ---  &     ---         & failed  & 58 \\
    LCG32, random &           32 &       3    & 14     &     ---         & passed  & 58 \\
    LCG64     &               64 &       None &  6     &     ---         & failed  & 46 \\
    LCG64, random &           64 &       None &  2     &     8           & passed  & 46 \\
    MWC       &          64 + 32 &       1    & 29     &     ---         & passed  & 44 \\
    Fibonacci, $r=521$ & $\ge 80$ &      None & 2      &     ---         & failed  & 23 \\
    Fibonacci, $r=1279$ & $\ge 80$ &     None & (1)    &     2           & passed  & 23 \\
    XORWOW (cuRAND) &          192 &       None & None   & 1/3             & failed & 19 \\
    MTGP (cuRAND) &       $\ge 44$ &       None &    2   &    2            & ---     & 18 \\
    XORShift/Weyl &           32 &       None & None   & None            & passed  & 18 \\
    Philox4x32\_7 &       (128)  &       None & None   & None            & passed  & 41 \\
    Philox4x32\_10 &      (128)  &       None & None   & None            & passed  & 30 \\ \hline
  \end{tabular}
\end{table}

\section{Counter-based generators\label{sec:counter}}

The essential complication of parallelizing random-number generators is rooted in the
fact that they are inherently recursive and thus appear to be, at first sight,
intrinsically serial. Also, it is this recursion that necessitates to store a
generator state in between invocations (and, therefore, uses up precious bandwidth
for loads and stores). The extremely simple Weyl generator discussed in the previous
section appears to be a notable exception as the one-step expression $y_n = (y_0 +
nc) \mod 2^w$ is of the form
\begin{equation}
  \label{eq:counter_based}
  x_n  = f_k(n),  
\end{equation}
that is, the $n$th number in the sequence is determined directly by applying some
function $f_k$ to the integer $n$ itself, which therefore acts as a {\em
  counter\/}. Here, $k$ is interpreted as a {\em key\/} representing all or part of
the parameters such as $y_0$, $c$ and $w$ for the Weyl sequence. Note that if $n$ is
a $w$-bit counter and $f_k$ is a bijection, the period of this type of generator is
$2^w$. While the Weyl generator itself certainly is not good enough as a RNG meeting
today's standards, one might wonder whether there are better generators based on the
same idea. A number of functions designed along these lines have been recently
discussed by Salmon {\em et al.\/} in Ref.~\cite{salmon:11}.

While such generators have not received much attention in the past for use in
simulations, they are very alike to the functions used in secret-key {\em
  cryptography\/}. There, one has a family of encryption functions that depend on a
key $k$ and encode the plain-text $n$ (represented as an integer) into a cipher-text
$x_n$ (encoded as another integer) \cite{trappe:book}. This is clearly of the form
\eqref{eq:counter_based}, assuming that $f_k$ is a bijection allowing to uniquely
decode the cipher-text. The connection to random-number generation comes in through
the security requirements of such function sets $f_k$ in cryptographic applications:
if the cipher-texts contain any structure that allow to distinguish them from pure
random sequences, this is a weakness of the system potentially allowing to break the
code (i.e., to find the key or plain-text only knowing the cipher-text). It is
therefore well-known in the cryptographic community that established systems such as
DES (the data encryption standard) and AES (the advanced encryption standard) can be
viewed as extremely high-quality random-number generators \cite{hellekalek:03}.

AES is an iterative block cipher based on the repeated application of keyed
bijections in several rounds designed to ensure diffusion of bits, i.e., generation
of highly random output from highly regular inputs. It uses a so-called
substitution-permutation network, which applies repeated substitutions (S-boxes) and
permutations (P-boxes) to the bits of the chosen block of the plain-text. The
resulting bijections are highly non-linear, in contrast to most of the
transformations used in traditional RNGs. The block size in AES is 128 bits, leading
to a more than sufficient period of $2^{128}\approx 3\times 10^{38}$. For details of
the AES transformation see, e.g., Ref.~\cite{trappe:book}. The authors of
Ref.~\cite{salmon:11} implemented AES as an RNG on CPU (there using the hardware
support for AES built into recent Intel and AMD CPUs) and GPU. These techniques
produce pseudo-random sequences passing all tests, but they are relatively slow
unless the mentioned special-purpose hardware support is available (which is not the
case on current GPUs).

To provide faster generators without hardware support, Salmon {\em et al.\/} suggest
a simplified schedule based on cryptographic techniques. The core component is based
on integer division and its remainder,
\[
\begin{split}
  \operatorname{mulhi}(a,b) &= \lfloor (a\times b)/2^w \rfloor,\\
  \operatorname{mullo}(a,b) &= (a\times b) \mod 2^w,
\end{split}
\]
which can be performed efficiently on most architectures (often reducing to one
machine instruction). The main iteration (or S-box) picks two words ($L,R)$ out of a
block of $N$ words of $w$ bits and computes
\[
\begin{split}
  L' &= \operatorname{mullo}(R,M),\\
  R' &= \operatorname{mulhi}(R,M)\oplus k \oplus L. 
\end{split}
\]
The final output is the result of $r$ rounds (so-called Feistel iterations) of the
application of $N/2$ such S-boxes with different multipliers $M$ but, for each
thread, constant ``key'' $k$. For $N > 2$, the $N$ elements are additionally permuted
in between rounds (P-boxes). Since multiplication, permutation and XOR $\oplus$ are
bijective, it is clear that the transformation is bijective. The quality of the
resulting class of generators, dubbed Philox-Nxw\_r, can be systematically improved
by increasing the number $r$ of Feistel iterations. The authors of
Ref.~\cite{salmon:11} empirically find that $r\ge 7$ is required for $N=4$ and $w=32$
to achieve Crush-resistance.

We tested the generators Philox-4x32\_7 and Philox-4x32\_10 on GPU. Note that,
although this generator requires local storage for four 32-bit words (128 bits) for
performing the iterations, it does not require to load or store a state, such that
its storage requirement is of a different nature than those of the recursive
generators. The use of four words leads to a period of $2^{128}$. Since such
(pseudo-)cryptographic bijections are designed to deliver outputs essentially
indistinguishable from random sequences for {\em any\/} choice of $k$, different keys
can be used to generate independent random-number streams in parallel
simulations. For the 64-bit key used here, this allows to generate $2^{64}$
independent random-number sequences, ideal for the parallel applications considered
here. Since the key and sequence space can be partitioned arbitrarily, it is
straightforward to use intrinsic logical variables to determine the random numbers to
be used. In a parallel Monte Carlo simulation, therefore, the counter could
correspond to an iteration or sweep number, whereas the key might be chosen to
represent the particle/spin number and further parameters (temperature, disorder
realization, system size, $\ldots$) characterizing the whole run. Using intrinsic
variables for sub-stream selection has the additional advantage of producing identical
results for any specific execution configuration on GPU or even between CPU and GPU
implementations.

The random-number streams of these generators have already been tested in various
sample-orders against the batteries in TestU01 in Ref.~\cite{salmon:11} and were
found to pass all tests. In addition to that, we used the Ising application test and
found no deviations, cf.\ the data in Table \ref{tab:rng}. Regarding execution speed,
we find the standalone generators to perform at $30\times 10^9$ for $r=10$ and
$41\times 10^9$ for $r=7$, which is better than the other high-quality generators
considered here. Similarly, in the Ising test the Philox performance is rather
good. Note that Philox4x32 produces four 32-bit random numbers per invocation. For
the Ising simulation with $k=1$, however, each thread only consumes two numbers such
that the $k=1$ performance results are, in fact, disfavoring Philox and could be
improved by rearranging the code accordingly. For $k=2$, were all produced random
numbers are also consumed, Philox4x32\_7 results in $0.1379$ ns per attempted spin
flip compared to $0.1278$ ns for the LCG32 and $0.1330$ for LCG64.

\section{Conclusions}

The generation of high-quality random numbers is an issue of continuing interest for
those engaging in computer simulation studies. After a number of unpleasant surprises
in the early years \cite{ferrenberg:92,parisi:95}, the community has today at its
disposal a number of generators with very long periods and passing most statistical
tests for simulations on serial machines such as single CPUs. With the advent of
massively parallel machines (some of the current clusters already have a total of
more than 1 million GPU cores), the search for adaptations of proved generators or
the invention of entirely new RNGs has begun. Apart from the more general problem of
parallel computing to provide an exceedingly large number of independent random-number
streams, simulations on GPU are faced with the additional challenge of finding
generators with small state or with the possibility of flexible state sharing between
the threads of a block (or warp) to accommodate the small amount of memory local to
the multiprocessors and the memory bandwidth limitations.

Simple, small-state generators such as linear congruential or multiply-with-carry
variants can be very fast. The statistical quality of the resulting sequences,
however, is not ideal for high-precision applications. State sharing, allowing
generators with larger states and hence much longer periods and ensuing better
statistical quality of produced numbers, appears to be a much more promising
strategy, providing some generators passing all tests of the extensive TestU01
suite. We suggest a new generator along these lines, based on the XORShift idea
proposed by Marsaglia \cite{marsaglia:03a}, which passes all tests and provides very
good performance. For some generators of this type, such as the Mersenne twister for
graphics processors suggested in Ref.~\cite{saito:10} and included in the latest
version of the cuRAND library, however, generation of appropriate parameter sets for
a large number of parallel instances is a challenging problem in itself. A completely
different approach \cite{salmon:11} based on the keyed bijections used in
symmetric-key cryptosystems is very versatile in producing large numbers of
independent random-number streams, does not require to save a state by coupling keys
and counters to intrinsic variables such as particle numbers and additionally
provides one of the most performant high-quality generators currently available on
GPUs.

\section*{Acknowledgments}

M.W.\ thanks W.\ Peterson for bringing Ref.~\cite{salmon:11} to his attention. M.W.\
acknowledges support by the DFG under contract No.\ WE4425/1-1 (Emmy Noether Programme).


\end{document}